\documentstyle[epsf]{elsart}

\begin{document}
\begin{frontmatter}
\title{Intermediate baseline appearance
experiments and
three-neutrino mixing schemes}

\author{Christian Y. Cardall and George M. Fuller}
\address{Department of Physics, University of California, San Diego,
La Jolla, CA 92093-0319, USA}
\author{David B. Cline}
\address{Physics \& Astronomy Department, Box~951547,
University
of California Los~Angeles,  Los Angeles, CA 90095-1547, USA}

\begin{abstract}
Three-neutrino mixing schemes suggested by Cardall \& Fuller
and Acker \& Pakvasa are compared and contrasted. Both of
these schemes seek to solve the solar and atmospheric neutrino
problems {\em and} to account for the possible neutrino oscillation signal
in the LSND experiment. These neutrino oscillation schemes have
different atmospheric and solar neutrino signatures that will be
discriminated by Super-Kamiokande and SNO. They will also have
different signatures in proposed long-baseline accelerator and 
reactor experiments. In particular, both of these schemes  
would give dramatic (and dramatically different) signals in an
``intermediate baseline'' experiment, such as the 
proposed ICARUS detector in the Jura mountains 
17 km from CERN.
\end{abstract}
\end{frontmatter}

\section{Three-mixing schemes}

At the present time there are three hints of neutrino
mixing: the solar (e.g. \cite{bahcall}) and 
atmospheric (e.g. \cite{atrev}) neutrino problems,
and the signal in the LSND experiment \cite{lsnd}. 
Each of these
can be solved by or interpreted as neutrino 
flavor oscillations. 
If an independent neutrino
mass difference
is associated with each of these three hints, 
four neutrino flavors are necessary. Since it is known
that only three neutrino flavors participate in the weak 
interaction \cite{zwidth}, the fourth neutrino flavor must be 
``sterile'' [an SU(2) singlet]. Accounting for all
three hints with only three active neutrino flavors
would require that at least
two of the phenomena must ``share'' one of the two
available independent mass differences. 

The possibility of a three-neutrino
mixing scheme along these lines was 
noted by
Cardall \& Fuller \cite{cardall96}. Their scheme uses the
smaller neutrino mass difference to account
for the solar neutrino problem, and the larger
mass difference to account for the atmospheric
neutrino problem and the LSND signal.
They proposed the following 
mass/mixing parameters:
\begin{eqnarray}
\delta m_{21}^{\ \ 2} &\approx& 7\times 10^{-6}\ \rm{eV}^2, \\
\delta m_{31}^{\ \ 2} &\approx& \delta m_{32}^{\ \ 2}
	\approx 0.3\ \rm{eV}^2, 
\end{eqnarray}
\begin{equation}
U_{\rm{CF,sma}} \approx \pmatrix{0.994 & 0.044 & 0.100 \cr -0.108 
	& 0.530 &
	0.841 \cr -0.015 & -0.847 & 0.532}.
\end{equation}
Here $\delta m_{ji}^{\ \ 2} = m_j^{\ 2}-m_i^{\ 2}$,
where $m_{i,j}^{\ \ 2}$ are the $i$th and $j$th 
squared neutrino mass eigenvalues, and we take
$ m_j^{\ 2} > m_i^{\ 2}$. The unitary matrix $U_{\rm{CF,sma}}$ 
(where ``CF'' stands for ``Cardall \& Fuller'' and
``sma'' stands for ``small mixing angle'') connects
the neutrino flavor and mass eigenstates: $\nu_\alpha =
\sum_i U_{\alpha i} \nu_i$, where $\alpha = e,\mu,$ or
$\tau$ and $i=1,2,3$. These mass/mixing parameters
were designed to employ the small mixing angle MSW solution
to the solar neutrino problem (see below). 
However,  as can readily be seen
from the analysis of Ref. \cite{threesol}, 
the general framework of the Cardall \& Fuller
scheme can also accomodate the solar neutrino
large mixing angle MSW solution:
\begin{eqnarray}
\delta m_{21}^{\ \ 2} &\approx& 2 \times 10^{-5}\ \rm{eV}^2, \\
\delta m_{31}^{\ \ 2} &\approx& \delta m_{32}^{\ \ 2}
	\approx 0.3\ \rm{eV}^2, 
\end{eqnarray}
\begin{equation}
U_{\rm{CF,lma}} \approx \pmatrix{0.873 & 0.478 & 0.100 \cr -0.330 
	& 0.428 &
	0.841 \cr 0.359 & -0.767 & 0.532},
\end{equation}
and the solar neutrino ``just-so'' vacuum oscillation solution:
\begin{eqnarray}
\delta m_{21}^{\ \ 2} &\approx& 6 \times 10^{-11}\ \rm{eV}^2, \\
\delta m_{31}^{\ \ 2} &\approx& \delta m_{32}^{\ \ 2}
	\approx 0.3\ \rm{eV}^2, 
\end{eqnarray}
\begin{equation}
U_{\rm{CF,vac}} \approx \pmatrix{0.807 & 0.582 & 0.100 \cr -0.381 
	& 0.384 &
	0.841 \cr 0.451 & -0.717 & 0.532}.
\end{equation}
Another variation on the Cardall \& Fuller scenario
has very recently been proposed in Ref. \cite{ma}.

A rather different scheme has been suggested recently
by Acker \& Pakvasa \cite{acker}, in which 
the smaller neutrino
mass difference accounts for both the solar and
atmospheric neutrino problems, and the larger mass 
difference accounts for the LSND signal. They
propose the following mass/mixing parameters:
\begin{eqnarray}
\delta m_{21}^{\ \ 2} &\approx& 10^{-2}\ \rm{eV}^2, \\
\delta m_{31}^{\ \ 2} &\approx& \delta m_{32}^{\ \ 2}
	\approx 1-2\ \rm{eV}^2, 
\end{eqnarray}
\begin{equation}
U_{\rm{AP}}\approx\pmatrix{0.700 & 0.700 & 0.140 \cr -0.714 & 0.689 &
	0.124 \cr -0.010 &-0.187 & 0.982}. 
\end{equation}
Relatively small alterations of this mixing matrix could also
allow this scheme to accomodate  slightly
larger or smaller mass differences to account for 
a significant fraction of the LSND
signal range.

\section{Consequences for solar neutrino experiments}

The solar neutrino problem arises from the observation that
the measured solar $\nu_e$ fluxes in experiments
with chlorine \cite{chlorine}, water \cite{water}, and gallium 
\cite{gallium} detectors are
only a fraction of the fluxes expected from models
of the sun. Furthermore, the fact that the detectors
employing the above substances have different thresholds
and that they observe different
$\nu_e$ flux deficits points to an energy dependence of
the $\nu_e$ suppression. This energy dependence picks
out values of $\delta m^2$ that may provide a neutrino
mixing resolution to the puzzle (e.g. 
\cite{threesol,solar}): $\delta m^2 \approx
10^{-5}\ \rm{eV}^2$ for an MSW effect (matter-enhanced
neutrino flavor conversion) solution, with both small and large
mixing angle solutions as possibilities;
or $\delta m^2 \approx
10^{-10}\ \rm{eV}^2$ for a ``just-so'' vacuum oscillation
solution, with large mixing angle.

While the Cardall \& Fuller schemes employ these
standard solar neutrino solutions,
the Acker \& Pakvasa scheme employs an ``energy independent''
solar neutrino solution with 
$\delta m^2 \approx 10^{-2}$ \cite{acker0,harr}. Their motivation
is to use the same mass difference as that suggested by the 
claimed zenith  angle dependence of the Kamiokande 
multi-GeV atmospheric
neutrino data (see below).  Such an energy independent solution
is strongly disfavored if all solar neutrino experiments and the
latest standard solar models are used \cite{krastev}, 
but is marginally possible if
the $^8$B neutrino flux is significantly (factor of $\sim 0.6$)
smaller than that predicted by the standard solar model. 
The possibility of a $^8$B neutrino flux smaller than that of
the SSM has received some motivation from a recent
measurement \cite{moto} of the reaction $\gamma + ^8$B $\rightarrow
^7$Be +p, the inverse of a reaction important to the
determination of that flux. However, issues surrounding 
the ability to extract the forward cross section from measurements
of the inverse reaction have been controversial \cite{coul}.
Another possibility for allowing a larger neutrino mass
difference than those in the ``standard'' solar neutrino solutions
is to ignore either the chlorine or gallium experiments
\cite{harr,krastev}.

The current evidence for neutrino oscillations associated
with the sun rests
simply on an observed neutrino flux deficit as compared 
with solar models.
Two next-generation solar neutrino experiments, SNO 
\cite{SNO} and
Super-Kamiokande \cite{SK}, will allow tests that can provide
more conclusive evidence of neutrino mixing. These
tests include time variable solar neutrino signals
(seanonal for vacuum oscillations \cite{foglivac,both}, day-night for 
the MSW solutions \cite{msw,both2}) and distortions of the recoil electrons
from  neutrino interactions in the detector \cite{both,both2,spec} 
(relevant for both the MSW solutions
and the ``just so'' vacuum oscillation solution).\footnote{SNO, with
its use of heavy water, will allow the very important measurement
of the ratio of charged current/neutral current  events for neutrino
interactions with deuterium.} These effects will be visible for
the Cardall \& Fuller 
schemes since they adopt the standard neutrino oscillation
solutions. However, these effects will be absent in 
the ``energy independent'' solar neutrino solution of 
Acker \& Pakvasa. 
This is a major observational difference between the Cardall \& Fuller 
and Acker \& Pakvasa mixing schemes.

\section{Consequences for atmospheric neutrino experiments}

Fluxes of $\nu_e$ and $\nu_\mu$ arise from
the interactions of cosmic rays with the earth's atmosphere. In
order to minimize uncertainties associated with neutrino
cross sections and a lack of
knowledge of the absolute magnitude of the neutrino fluxes, 
the ``ratio of ratios'' $R$ is a commonly reported observable:
\begin{equation}
R = { {(\nu_{\mu} / \nu_e)_{\rm{data}} } \over 
  {(\nu_{\mu} / \nu_e)_{\rm{Monte \; Carlo}} } }.
\end{equation}
Studies involving large water detectors (Kamiokande 
\cite{atmkam} and IMB \cite{atmimb})
have reported  values of $R\sim 0.6$.
Experiments using iron calorimeter detectors have 
had mixed results: Soudan 2  \cite{soudan} has observed a deficit in $R$
comparable to that seen in the water detectors,
while NUSEX \cite{nusex} and Fr\'ejus \cite{frejus} 
have not seen evidence for 
values of $R$ significantly different from
unity. 

An interesting aspect of the atmospheric neutrino puzzle is
the zenith-angle dependence of $R$ 
reported by the 
Kamiokande group for multi-GeV neutrinos \cite{zenith}.  
According to their analysis, $R$ decreases
from zero zenith angle (neutrinos from the atmosphere immediately
above the detector) to maximum zenith angle (neutrinos from
the atmosphere on the other side of the earth). This dependence
suggests that neutrinos from overhead do not travel far enough
to oscillate significantly, while those coming from across the earth
do travel sufficiently far to oscillate. Such a scenario restricts 
neutrino mass differences to the range $10^{-3} - 10^{-1}\ 
\rm{eV}^2$.

A neutrino mass difference of about $10^{-2}\ \rm{eV}^2$ has
therefore come to be associated with the atmospheric neutrino
problem. This is the mass difference employed in the Acker
\& Pakvasa neutrino oscillation scheme. The mass difference
of $\sim 0.3\ \rm{eV}^2$ in the Cardall \& Fuller scheme requires
one to discount the zenith angle dependence of the Kamiokande
multi-GeV data. This position is not entirely implausible, as
the statistical significance of this effect has been questioned
\cite{stat},
and the IMB group saw no zenith angle dependence in their data
\cite{clark}.

Super-Kamiokande will be able to settle the question of 
zenith angle dependence definitively. Preliminary analyses
already confirm a deficit in $R$ averaged over all directions,
but nothing definitive can yet be said about the zenith angle
dependence \cite{haines}. 
This is perhaps interesting in view of the fact that 
Super-Kamiokande already has more events than were
obtained with the previous Kamiokande detector.

Another important point is that in the
Acker \& Pakvasa
solution
the atmospheric neutrino oscillations are primarily
$\nu_\mu \leftrightarrow \nu_e$  to 
allow a simultaneous solution to the solar neutrino
problem, while
in the Cardall \& Fuller scheme the atmospheric 
neutrino oscillation channel is primarily
$\nu_\mu \leftrightarrow \nu_\tau$ to avoid
conflict with $\nu_e$ disappearance experiments. 
Due to uncertainties in the $\nu_e$ and
$\nu_\mu$ absolute fluxes, it is not currently possible
to determine the oscillation channel of atmospheric
neutrinos. However, higher precision cosmic ray measurements
may help reduce the
ambiguity.

\section{Consequences for accelerator and reactor experiments}

While solar and atmospheric neutrinos have provided tantalizing
suggestions of neutrino mixing, it is highly desirable
to directly observe neutrino oscillations in 
controlled terrestrial experiments. 
We point
out that the neutrino mass difference associated with solar neutrinos
is so small that all three of the Cardall \& Fuller solutions,
corresponding to different solar neutrino solutions, will have the
same effects in the terrestrial experiments discussed here.

\subsection{Short baseline experiments}
The LSND experiment, which has reported  an excess
of ${\bar\nu}_e$ in a ${\bar\nu}_\mu$ beam from pion
decays at rest, continues to run. KARMEN \cite{karmen}, 
a similar experiment,
has recently received an upgrade that should eventually be
able to confirm this LSND signal. While the LSND signal
is compatible with large neutrino mass differences, negative
results from CCFR \cite{ccfr}, CHORUS \cite{chorus}, and 
NOMAD \cite{nomad} restrict the viable
range of of neutrino mass difference to $\delta m^2 \le
10\ \rm{eV}^2$.
In addition, LSND has recently completed a study 
of excess $\nu_e$
events in a $\nu_\mu$ beam from pion decays in flight
\cite{flight},
a channel whose backgrounds and systematics
are different from those of the ${\bar\nu}_e$ beam from
pion decays at rest.
This excess can also be accounted for by neutrino oscillations,
with mixing parameters that overlap those suggested by
the decay at rest data. 
Finally, COSMOS \cite{cosmos},
the proposed short baseline counterpart to the MINOS
experiment, should just be able to see $\nu_\mu \leftrightarrow
\nu_\tau$ oscillations in both the Acker \& Pakvasa and
Cardall \& Fuller schemes.

\subsection{Long baseline experiments}
Long baseline experiments are designed to explore
the region of neutrino mixing parameter space suggested by
the atmospheric neutrino problem, neutrino mass differences of
 $\sim 10^{-2}\ \rm{eV}^2$ in particular. Long baseline accelerator
experiments such as MINOS (Fermilab to the Soudan mine)
\cite{minos},
KEK to Super-Kamiokande \cite{kek}, and 
CERN to Gran Sasso \cite{cerngs} involve 
a $\nu_\mu$ beam and should be able to distinguish between
$\nu_\mu \leftrightarrow \nu_\tau$ and $\nu_\mu \leftrightarrow
\nu_e$ oscillations. Figure 1 shows the oscillation
(flavor conversion) probabilities
for these channels implied by the Cardall \& Fuller and Acker
\& Pakvasa schemes for a baseline of 735 km, which coincidentally
is the approximate baseline length for both the CERN to Gran Sasso
and MINOS experiments.
The differences are evident. The Acker \&
Pakvasa scheme has large amplitude $\nu_\mu \leftrightarrow
\nu_e$ mixing, with oscillations in energy space that could probably
be observed experimentally. The Cardall
and Fuller scheme has large amplitude $\nu_\mu \leftrightarrow \nu_\tau$
mixing, with the oscillations in energy space being so rapid that in
practice an average would probably be measured. 

There are also forthcoming long baseline reactor experiments,
CHOOZ \cite{chooz} and Palo Verde \cite{palo}. 
These are ${\bar\nu}_e$ disappearance experiments
which will be sensitive to neutrino mass differences as low as
$\sim 10^{-3}\ \rm{eV}^2$, but which will not provide increased sensitivity
to the oscillation amplitude. These experiments are therefore sensitive
to a $\nu_\mu \leftrightarrow
\nu_e$ solution to the atmospheric neutrino problem
(which has a very large effective mixing angle). They {\em would}
see neutrino oscillations predicted by the Acker \& Pakvasa scheme,
but {\em not} those predicted by the Cardall \& Fuller solution.


\begin{figure}
\epsfxsize=12cm \epsfbox{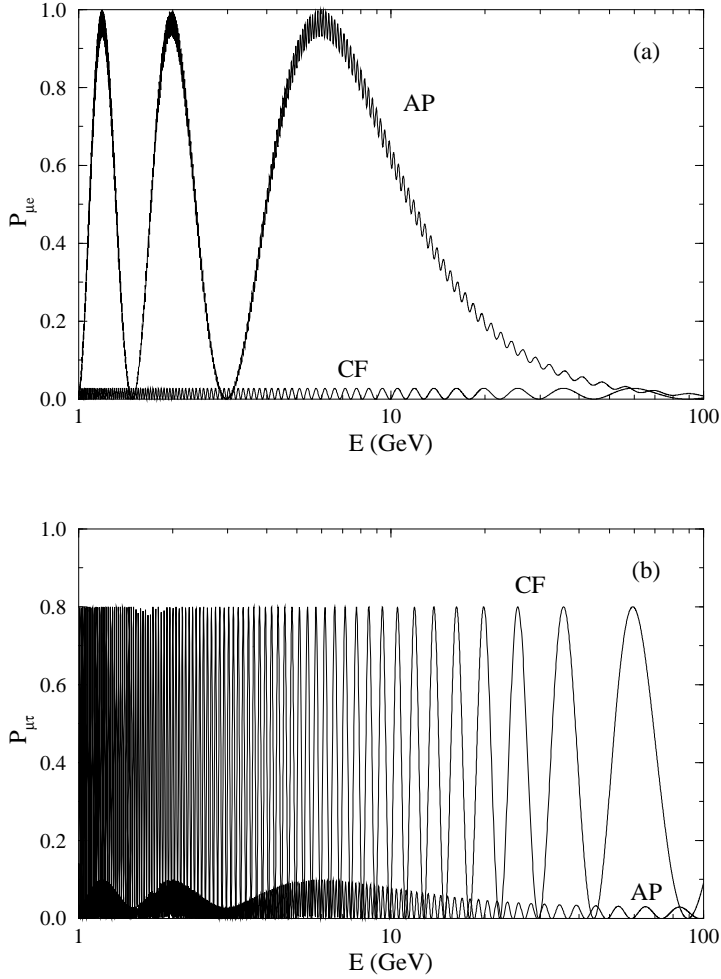}
\caption{Probabilities for (a) $\nu_e$ and (b) $\nu_\tau$
	appearance in a $\nu_\mu$ beam, as a function of
	neutrino energy, for a baseline of 735 km (CERN to
	Gran Sasso and Fermilab to the Soudan mine). ``CF''
	and ``AP'' stand for the Cardall \& Fuller and
	Acker \& Pakvasa oscillation schemes respectively.}
\end{figure}

\begin{figure}
\epsfxsize=12cm \epsfbox{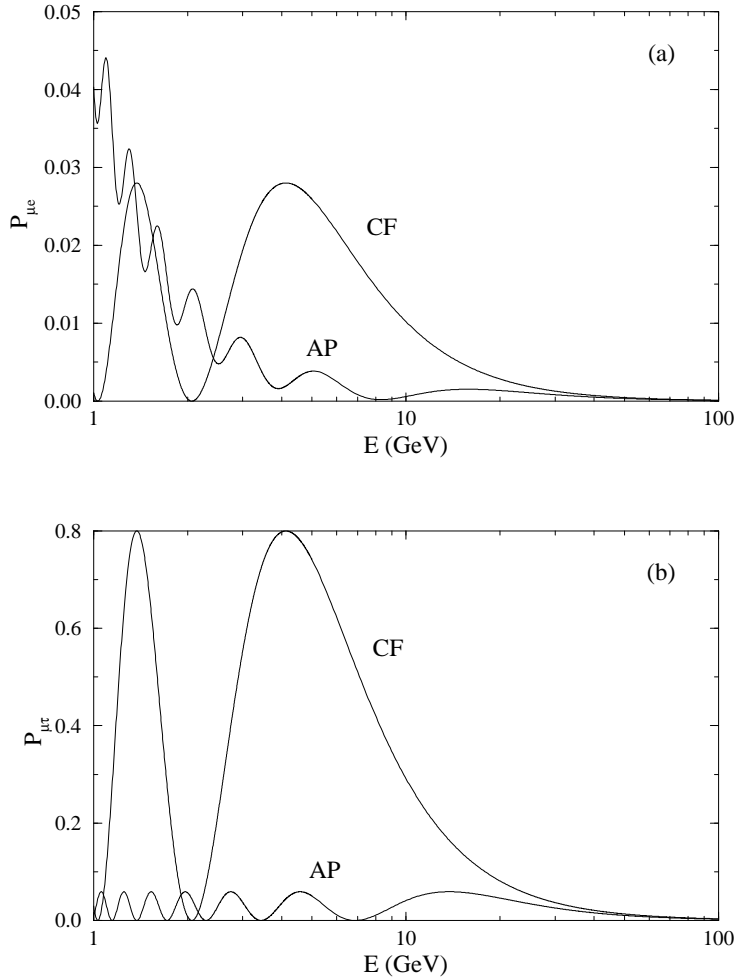}
\caption{Probabilities for (a) $\nu_e$ and (b) $\nu_\tau$
	appearance in a $\nu_\mu$ beam, as a function of
	neutrino energy, for a baseline of 17 km (CERN to
	the Jura mountains). ``CF''
	and ``AP'' stand for the Cardall \& Fuller and
	Acker \& Pakvasa oscillation schemes respectively.}
\end{figure}

\subsection{The Jura mountain neutrino detector}
The ICARUS collaboration, which is producing neutrino detectors
for the CERN to Gran Sasso long baseline experiment,
 has recently proposed placing
one of their detector modules behind the Jura mountains
17 km from CERN \cite{jura}. This ``intermediate
 baseline'' experiment is
of great interest in light of the three-neutrino mixing schemes
discussed here, since the neutrino flight distance divided by
the neutrino oscillation length is such that neutrino oscillations should
be directly observable in energy space, as illustrated in Figure 2
[the estimated energy resolution for ICARUS is (5-15)\%].
This experiment was proposed with the Acker \& Pakvasa scheme
in mind, but Figure 2 shows that the Cardall \& Fuller scheme
predicts an even more spectacular signal in the $\nu_\tau$
appearance channel. We emphasize that an intermediate baseline 
experiment such as this is important to distinguish between
the Acker \& Pakvasa and Cardall \& Fuller schemes, and
to directly observe two features of those schemes which
are presently only inferred from phenomenological considerations:
(1) a neutrino mass difference in the range $0.1 \le \delta m^2 \le 1$
eV$^2$, and (2) significant $\nu_\mu \leftrightarrow \nu_\tau$ 
mixing, which is (at least in part) indirectly responsible for the
 $\nu_\mu \leftrightarrow \nu_e$ signal observed at LSND.

\begin{table}
\caption{Some experimental signatures of the 
	Cardall \& Fuller and Acker \& Pakvasa schemes.}
\begin{tabular}{lcc}
\hline
Experimental signature & Cardall \& Fuller & Acker \& Pakvasa \\
\hline
{\em Solar neutrinos:} \\
\ \ \ Energy dependence &  Yes & No \\
\ \ \ Spectral distortions & Yes & No \\
\ \ \ Time dependence	& Yes & No \\
{\em Atmospheric neutrinos:} \\
\ \ \ Zenith angle variation & No & Yes \\
{\em Long baseline reactor:} \\
\ \ \ ${\bar\nu}_e$ disappearance & No & Yes \\
{\em Long baseline accelerator:} \\
\ \ \ $\nu_e$ appearance & No & Yes (unaveraged) \\
\ \ \ $\nu_\tau$ appearance & Yes (averaged) & No \\
{\em Intermediate baseline accelerator:} \\
\ \ \ $\nu_e$ appearance & No & No \\
\ \ \ $\nu_\tau$ appearance & Yes (unaveraged)& Yes (unaveraged)\\ 
\hline
\\ 
\end{tabular}
\end{table}

\section{Conclusion}

There exist at least two three-neutrino mixing 
schemes designed to satisfy three hints of neutrino mixing
(the solar and atmospheric neutrino problems, and the
signal at LSND). Some of their signatures in future
experiments are given in Table 1.
Accomodating all three of these hints
would normally require four neutrino flavors; 
use of a three generation framework requires ignoring
some aspect of the data. Cardall \& Fuller have chosen
to ignore the zenith angle dependence of the atmospheric
neutrino data, while Acker \& Pakvasa have chosen to
ignore the energy dependence of the solar neutrino data.
The former choice has recently been identified in a
thorough analysis as the
choice of ``minimum sacrifice'' \cite{minsac}, 
and the validity of the
above assumptions regarding atmospheric and solar
neutrinos will eventually be tested by Super-Kamiokande
and SNO.
Nevertheless, direct
experimental verification of the predictions of these
schemes in a controlled terrestrial experiment is highly
desirable. In particular, an ``intermediate baseline'' experiment
such as the proposed ICARUS detector to be placed
behind the Jura mountains 17 km from CERN is important
to be able to see, in energy space, unaveraged 
$\nu_\mu \leftrightarrow \nu_\tau$ oscillations driven by
a neutrino mass difference of $0.1 \le \delta m^2 \le 1$
eV$^2$. 

\begin{ack}
We thank M. Goodman and W. Haxton for helpful 
communications.
This work was supported 
by grant NSF PHY95-03384
\end{ack}

\end{document}